# Giant electro-optic coefficient in single-crystal barium titanate-on-oxide insulator based Mach-Zehnder interferometer

Hong-Lin Lin[1,2], Pragati Aashna[1,2], Yu Cao[1,2]* , Aaron Danner[1]*

[1]Department of Electrical and Computer Engineering, National University of Singapore, Singapore, Singapore

[2]These authors contributed equally: Hong-Lin Lin, Pragati Aashna, Yu Cao.

*Corresponding author e-mail: caoyu@uestc.edu.cn ; adanner@nus.edu.sg

**Electro-optic modulators are indispensable components of modern day photonic integrated circuits (PICs). Recently lithium niobate has emerged as a key material to realize large-bandwidth high-speed modulation, but next-generation modulators require high-density integration, low cost, low power and high performance simultaneously, which are difficult to achieve with established integrated lithium niobate photonics platforms due to its limited electro-optic coefficient. Leveraging its exceptional Pockels coefficient, barium titanate (BTO) in the thin film form has emerged as a promising alternative but the electro-optic coefficients reported in thin-film BTO often fall short of bulk values due to challenges in film growth and waveguide fabrication. Here, we report, to the best of our knowledge, the largest Pockels coefficient ($r_{42}$) of 1268 pm/V in thin film BTO platform, which is very close to the bulk value. We measure it by using an unbalanced Mach-Zehnder interferometer, fabricated by an optimized wet-etching method for realising single-mode waveguides in single-crystal barium titanate-on-insulator grown by pulsed laser deposition. This giant $r_{42}$ is extracted from a device in which the optical mode is fully confined within a single-crystal BTO waveguide. This approach contrasts with previous designs where the core material — typically silicon or silicon nitride — supports only partial mode confinement, resulting in an evanescent overlap with a multi-crystalline BTO layer. Our highly confined BTO-on-insulator with a strong electro-optic effect may significantly advance the field of ultra-low-power integrated photonic devices and allow for the realization of next-generation efficient and compact photonic circuits.**

Photonic integrated circuits (PICs) have emerged as drivers for the beyond-Moore technology, overcoming limitations of traditional copper-based systems by leveraging photons for data transmission at the speed of light[1]. Over the years, silicon photonics has matured as an integrated photonics platform, benefiting from microelectronics processing techniques to achieve high-speed, low-power data transmission. While electro-optic modulation is essential for on-chip light control, silicon's centrosymmetric crystal structure prevents a meaningful electro-optic effect, motivating alternatives like plasma dispersion, thermo-optics, or strain engineering[2–5]. Recently, thin-film nonlinear materials like thin-film lithium niobate on insulator (LNOI) have emerged as efficient, high-speed, and energy-optimized alternative photonic platforms owing to the large electro-optic (Pockels) effect availabe ($r_{33}$=30 pm/V in lithium niobate)[6]. The recent commercialization of thin-film LNOI and advances in device fabrication have enabled strong optical mode confinement, allowing for smal electrode gaps and hence strong electric fields, leading to efficient exploitation of the Pockels coefficient. These innovations significantly reduce the $V_{\pi}L$ product in traditional Mach-Zehnder interferometer (MZI) configurations, resulting in smaller device footprints, lower operating voltages, and higher bandwidths compared to conventional bulk lithium niobate modulators[7,8].

In the context of high Pockels effect materials, bulk single crystal barium titanate (BTO, or $BaTiO_3$) has excellent electro-optic properties ($r_{51}$=$r_{42}$=1300 pm/V)[9], over one order of magnitude stronger than those of lithium niobate, making it an attractive material for potential fabrication of nonlinear optical devices. Bulk BTO single crystals are costly and challenging to process due to their low Curie temperature, however, which limits their ability to withstand high fabrication temperatures and achieve high index contrast without fracturing. In contrast, thin-film lithium niobate has demonstrated significantly improved optical confinement and performance over its bulk form, underscoring the potential of thin-film approaches. Consequently, there is a strong impetus to advance research into thin-film BTO for parallel gains.

Following early exploration of thin-film BTO on MgO substrate[10,11], various epitaxial growth techniques have been developed to deposit high-quality BTO on substrates like silicon and dysprosium scandate (DSO)[12,13], achieving exceptionally large Pockels coefficients and avoiding fracturing due to a substrate-mediated increase in Curie temperature. Several techniques have consequently been explored to fabricate BTO photonic devices with methods like direct wet etching or dry etching, heterogenous integration with platforms like silicon nitride or silicon, as well as by engineering the local refractive index through different methods[10,12–14]. Among these methods, etching BTO thin films to make ridge waveguides is most desirable as this permits a strong optical mode confinement within the active electro-optic BTO layer, i.e. the optical overlap. However, most reported devices so far have relied instead on heterogeneous integration of BTO with silicon or silicon nitride, which allows for only partial evanescent interaction of propagating light with a BTO active layer. Therefore, despite a large BTO Pockels coefficient, actual device performance reports have been modest.

In this work, we report the extraction and exploitation of a large Pockels coefficient $r_{42}$ of 1268 pm/V from a BTO ridge waveguide etched such that a large optical mode overlap of 74.3% is simultaneously achieved. To measure the $r_{42}$, an unbalanced Mach-Zehnder interferometer (MZI) device based on thin film barium titanate-on-insulator (BTO/DSO) is fabricated with a wet-eching method[15]. This establishes that strong optical mode confinement within highly transparent BTO with a strong electro-optic effect can simultaneously be achieved without degradation of crystal quality or any decrease in effective Pockels parameters, and illustrates the potential of thin film BTO as a robust physical platform for a wide range of nonlinear optical applications, particularly in the development of large-bandwidth and high-efficiency modulators. The integration of thin film barium titanate (BTO) with advanced waveguide structures offers significant advantages in modulation speed and power consumption, facilitating the realization of next-generation modulators.

## Device modelling and design

We first grow a 500 nm thick BTO film on a 100 $mm^2$ DSO substrate by pulsed laser deposition, using a KrF laser (248 nm wavelength); growth details can be found in our earlier work[13]. We verify that the BTO is single-crystal with the c-axis oriented out of plane. This arrangement allows $r_{51}$ and $r_{42}$ to be exploited with the use of co-planar electrodes as shown in Fig. 1a.

COMSOL and FDTD simulations were used to design and optimize the waveguide configuration, aiming to achieve single-mode waveguides with compact bending radii. While ridge waveguides (fully etched) offer advantages in minimizing bending radii compared to rib waveguides (partially etched), they are prone to increased scattering loss due to sidewall roughness. Therefore, based on prior findings[15], we chose a 250 nm etch depth within a 500 nm thick BTO film on DSO. COMSOL analysis indicates that waveguides with a base width below 2.5 μm can support only the fundamental TE and TM modes, so this width was chosen to pattern our waveguides. Waveguides were wet-etched with 0.03% HF[15].

An unbalanced MZI with an arm length difference of 430 μm was fabricated as shown in Fig. 1. The electrode separation distance of 8 μm (2.75 μm away from waveguide edge on both sides) was optimized through simulation to ensure maximum overlap between the electric field and optical mode while avoiding propagation loss resulting from mode overlap with the metal of the electrode; the electrode length was 3500 μm. Root mean square (RMS) roughness of the etched surface was 2.52 nm, as shown in the atomic force microscope (AFM) image (Fig. 1b). Scanning electron microscope (SEM) images of the waveguides fabricated are shown in figure 1c-d. The etch angle was 45° as is typical for wet etching.

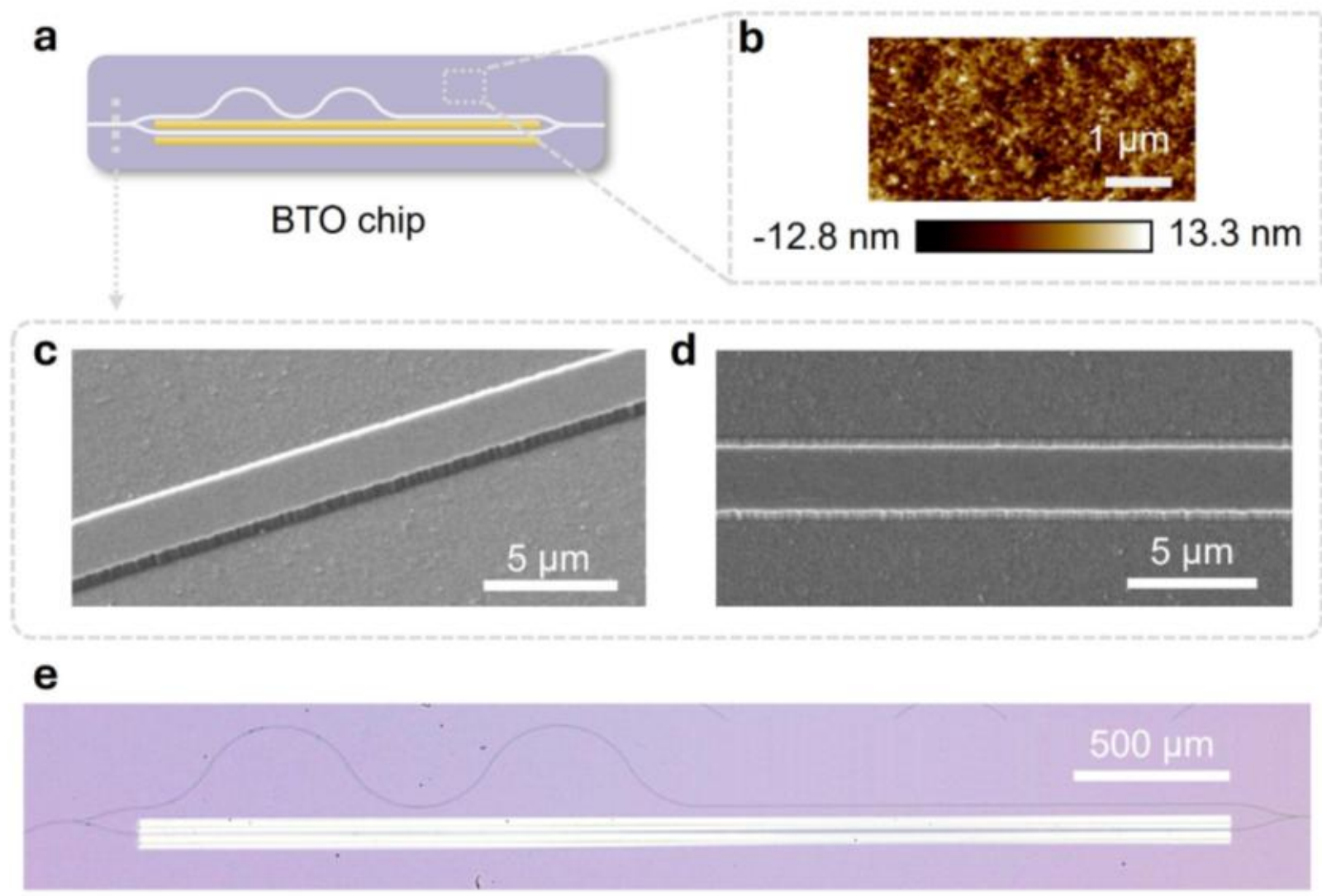

**Fig. 1 | BTO MZI device fabricated by wet etching.** **a**, A schematic drawing of the BTO MZI device. **b**, Atomic force microscopy (AFM) image of the surface morphology of the etched area as labelled in **a**, with a root mean square (RMS) roughness of 2.52 nm, measured after etching. **c**,**d** Scanning electron microscope (SEM) images of the etched BTO ridge; 45° angle view (**b**); top view (**c**). **e**, A full view optical image of the unbalanced MZI device. The device has an electrode length of 3500 μm and an arm length difference of 430 μm.

### Device characterization and electro-optic effect

MZI devices were experimentally characterized with the setup illustrated in Fig. 2a. Fig. 2b shows a simulated cross-sectional optical mode and electric field distribution, clearly illustrating strong mode confinement at these waveguide dimensions. TE polarized light is coupled into the MZI, and the measured transmission spectrum for the unbalanced MZI is shown in Fig. 2c. We increase the voltage in steps of 5 V to show a consistent shift in the transmission spectrum with an increase in applied voltage.

From the transmission spectrum, we observe that the free spectral range (FSR) of the transmission spectrum is 2.563 nm. With an applied voltage of 5 V, the resonance shift is 0.288 nm. We measure the resonance shift for voltages from 0 V to 20 V and we plot it as a function of applied voltage as shown in Fig. 2d. The plot indicates a linear relationship, and the resonance shift efficiency (or tuning efficiency) of 41.3 pm/V is extracted.

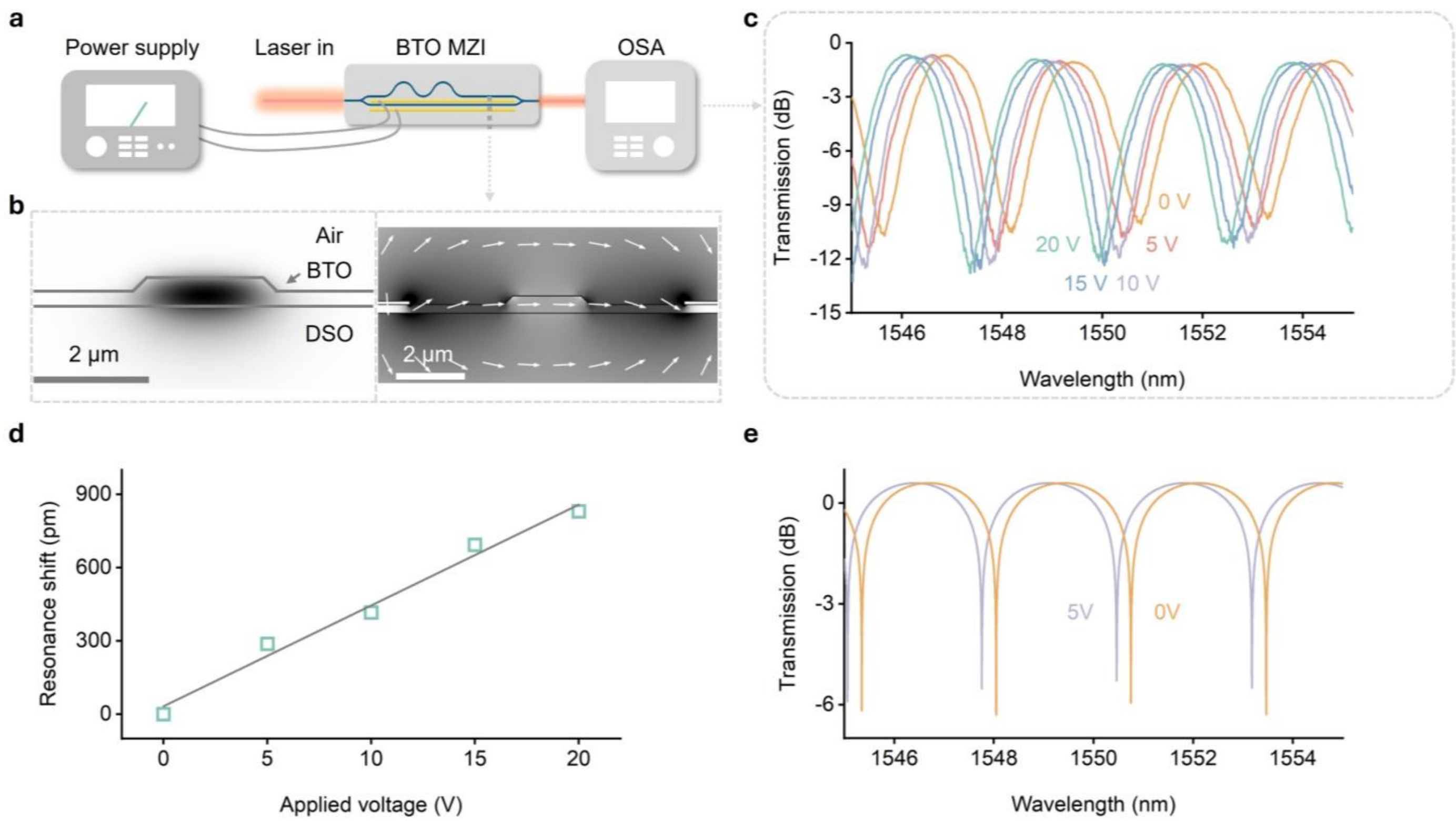

**Fig. 2 | Characterization of the electro-optic tuning of the BTO MZI devices. a**, Schematic representation of the measurement set-up. The linearly polarized light is coupled into the MZI device and the output is coupled to a fiber optic cable which is connected to an optical spectrum analyzer (OSA). **b**, Left: Cross-sectional mode analysis of the ridge waveguide (width of 2.5 µm, height of 250 nm). Fundamental TE mode with 74.3% optical power confined in active layer of BTO. The effective mode index is 2.066. Right: Simulated electric field distribution. 5 V is applied over an 8 µm gap, with an effective 3.49 kV/cm field strength at the BTO ridge area where the optical mode is located. **c**, Measured transmission spectrum with different applied external voltages. **d**, Resonance shift as a function of applied voltage, which represents a linear relation and tuning efficiency of 41.3 pm/V. **e**, Theoretically simulated transmission spectra for the geometry of the measured device; the yellow curve is at 0 V, and the purple curve is at 5 V.

The Pockels coefficient responsible for the observed electro-optic tuning of the MZI resonance can be extracted from the observed resonance shift. A trial value for $r_{42}$ can be used to calculate the BTO's (material) permittivity matrix at 5 V, which can then be used to create a cross-sectional analysis as shown in Fig. 2b and obtain mode parameters; propagation and interference in the MZI then lead to the (lossless) curves in Fig. 2e. By comparing the theoretical resonance shift at various $r_{42}$ values with the experimentally measured shift, an $r_{42}$ value of 1268 pm/V is found. Details are discussed in the Supplementary Material. (The procedure is not straightforward due to the fact that $r_{42}$ influences the BTO's off-diagonal permittivity tensor elements.)

In traditional electro-optic modulator design, modulation performance is primarily determined by the Pockels coefficient. The modulation efficiency, represented by the voltage-length product $V_{\pi}L$, is inversely proportional to it. Bandwidth limitations often arise from velocity mismatches between microwave and optical waves, which can be mitigated by using shorter electrodes. Consequently, the Pockels coefficient $r$ is a critical parameter. However, even with a high $r$ value and impressive device performance, in any hypothetical device with only limited optical overlap with the active BTO layer, the strong Pockels coefficient will be only partially utilized, reducing the overall electro-optic interaction[12,16]. This motivates increasing optical confinement as much as possible.

Fig. 3a and 3b illustrate measured Pockels coefficients in recently reported BTO devices; while reasonably large Pockels coefficients have been demonstrated, these were often limited by the thickness of the films and etch depth of the ridge waveguides (Fig. 3a), which in turn restricted the achievable mode confinement factor $\Gamma_{BTO}$ (Fig. 3b). Our current work simultaneously attains high values for both these parameters which can be attributed to good quality crystal as well as well-patterned waveguides. The optical overlap ( $\Gamma_{BTO}$) is influenced not only by the thickness of the BTO thin film (illustrated in Fig. 3c) but also by the design and fabrication of the waveguide (shown in Fig. 3d-f).

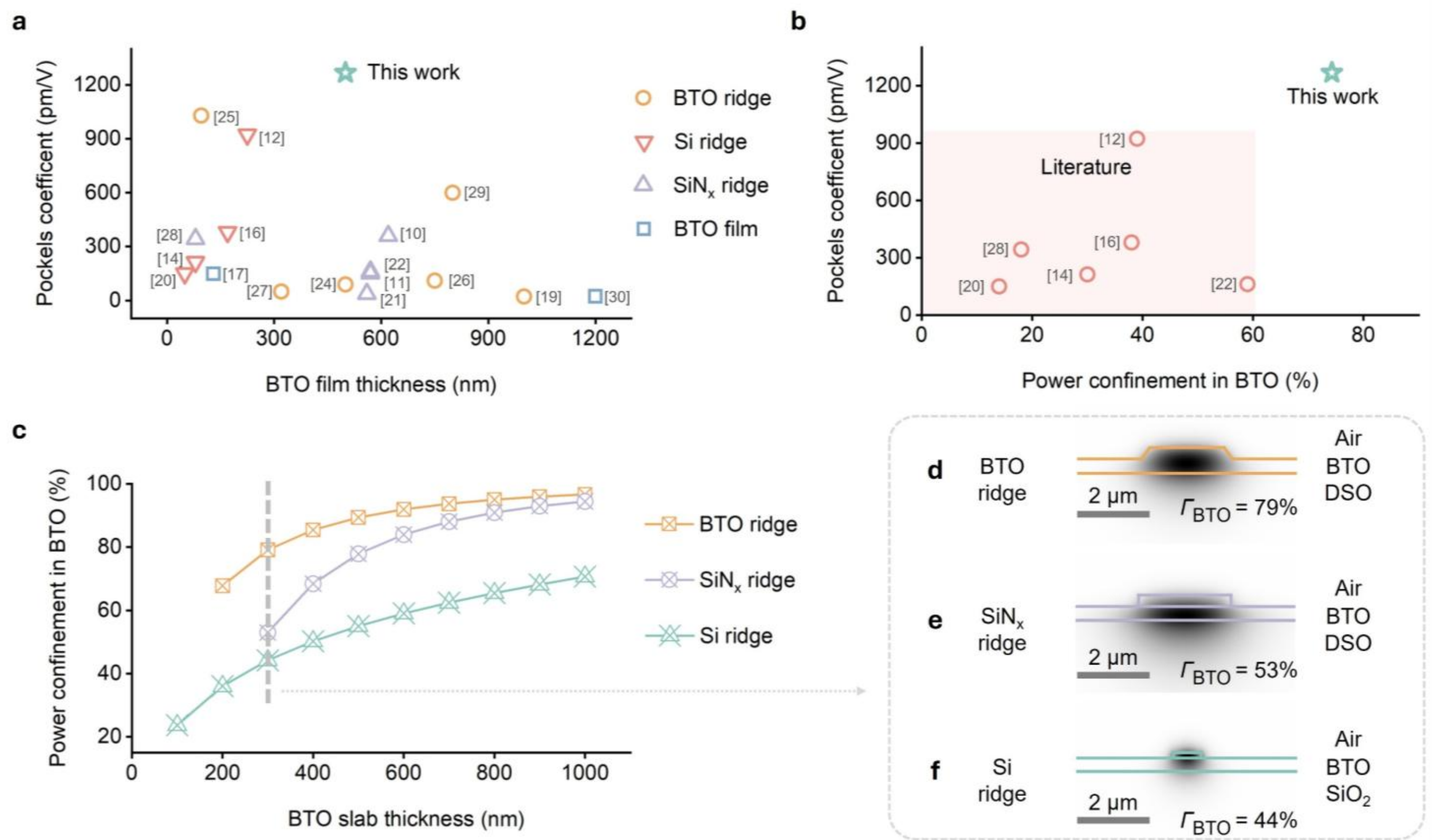

**Fig. 3 | Comparative analyses of the Pockels coefficient achieved in BTO thin films. a**, Comparative illustration of the Pockels coefficient achieved in BTO thin films and the corresponding film thickness. **b**, Comparative illustration of the Pockels coefficient achieved in waveguide structures (either BTO ridge waveguide, silicon nitride ridge waveguide, or silicon ridge waveguide, with the corresponding power confinement in the BTO layer). **c**, Power confinement factor in the BTO layer as a function of BTO slab thickness for BTO ridge, silicon nitride ridge and silicon ridge waveguide, respextively, simulated in COMSOL. The same geometry is used for BTO ridge and silicon nitride ridge for convenience of comparison, because their refractive indices are close to each other, a 250 nm ridge height and a 2.5 μm ridge width were used, the slab layer is BTO and the substrate cladding layer is DSO. The silicon ridge has a 90 nm ridge height and a 0.75 μm ridge width, with a BTO slab on $SiO_2$ substrate, as fabricated in reference 12. The dashed line labels the BTO slab thickness at 300 nm, where we later make comparison in **d-f**. **d-f**, Illustration of cross-sectional mode analyses for BTO ridge waveguide, silicon nitride ridge waveguide, and silicon ridge waveguide, respectively. [10–12,14,16–31]

**Table 1 | Comparison of various fabricated BTO waveguide devices with either large Pockels coefficients or large $\Gamma_{\mathrm{BTO}}$.**

| No. | Reference | Waveguide type | Pockels coefficient (pm/V) | $\Gamma_{\mathrm{BTO}}$ | Cartoon illustration of mode profile [a] |
|---|---|---|---|---|---|
| 1 | [28] | SiN ridge | $r_{\mathrm{eff}}$ = 343 | 18 % [b] | |
| 2 | [16] | Si ridge | $r_{\mathrm{eff}}$ = 380 | 38 % [b] | |
| 3 | [18] | Si-BTO-Si | $r_{\mathrm{eff}}$ = 624 | 11.66 % [b] | |
| 4 | [12] | Si ridge | $r_{42}$ = 923 | 39 % [b] | |
| 5 | [29] | BTO ridge | $r_{42}$ = 600 | 91.25% [c] | |
| 6 | This work | BTO ridge | $r_{42}$ = 1268 | 74.3 % | |
| 7 | [10] | SiN ridge | $r_{\mathrm{eff}}$ = 420 | 92.86% [c] | |
| 8 | [14] | Si-BTO-Si | $r_{\mathrm{eff}}$ = 213 | 30% [b] | |
| 9 | [19] | BTO ridge | $r_{\mathrm{eff}}$ = 22 | 97.9% [c] | |

[a] The cartoon was drawn by the authors to illustrate mode profiles and mode confinement for different structures to be commensurate with the original data from the respective references.

[b] The power confinement in BTO ($\Gamma_{\mathrm{BTO}}$) presented in the table is directly described in the reference.

[c] The power confinement in BTO ($\Gamma_{\mathrm{BTO}}$) presented in the table is estimated by COMSOL simulation with the parameters from the corresponding reference.

To highlight the significance of our findings, we compare BTO-based structures that have achieved either large Pockels coefficient or large $\Gamma_{\mathrm{BTO}}$ in Table 1. It is important to note that achieving such large Pockels coefficient and optical overlap at the same time requires an appropriately designed electro-optic platform and enhanced direct etching techniques, which are the factors leading to our current results[13].

Our data clearly shows a thin-film Pockels coefficient approaching the previously unattainable bulk value, while simultaneously demonstrating substantial overlap between electric and optical fields, all with a simple etched device structure which avoids the evanescent BTO field interactions with underpinned many previous reports of strong Pockels effects in BTO. This result could potentially lead unprecedented enhancement in future device performance, when considering the strength of the Pockels parameter relative to that of lithium niobate, for instance. The implications of these findings open exciting avenues for research aimed at developing high-speed, low-power, and compact modulators and other nonlinear optical devices, laying the foundation for the advancement of state-of-the-art performance in next-generation photonic integrated circuits (PICs) in BTO.

In summary, we have successfully grown high-quality 500 nm thin-film BTO on a DSO substrate and fabricated unbalanced MZI devices through wet etching, achieving an etching depth of 250 nm and a waveguide width of 2.5 µm, with an arm length difference of 430 µm. The transmission spectrum of the MZI demonstrates a free spectral range of 2.563 nm. Upon applying an external voltage to the MZI device, we observed a resonance shift of 41.3 pm/V, which exhibits a linear relationship with the applied voltage, and corresponds to a Pockels coefficient ($r_{42}$) of 1268 pm/V. This represents the largest yet-reported Pockels coefficient for thin film BTO, and combines this result with a large in-device optical overlap of 74.3%, laying the groundwork for the development of more compact nonlinear optical devices in integrated optics.

## Acknowledgement

The authors acknowledge support by the National Research Foundation (NRF), Singapore, under its Competitive Research Program (CRP Award No. NRF-CRP24-2020-0003) and A*STAR under its NSTIC programme, Project No. M24W1NS003.

## Author contributions

Y.C. and H.L. conceived this work. Y.C. grew the BTO sample. H.L. and P.A. designed the device. H.L. fabricated the device. H.L. designed and built the experimental set-up. H.L. and Y.C. performed the optical characterization. Y.C. performed the theoretical calculations and analyzed the data. H.L. performed the SEM and OM imaging. P.A. performed the AFM. Y.C. and H.L. performed the cross-sectional simulations. Y.C. performed visualization. Y.C. wrote the manuscript. Y.C., H.L, P.A. and A.D. revised the manuscript. A.D. supervised this work.

## Competing interests

The authors declare no competing interests.

## Methods

### BTO thin film growth

BTO films were grown by pulsed laser deposition with a KrF laser (248 nm wavelength). Laser energy density was fixed at 1.5 J $cm^{-2}$ and the repetition rate was 5 Hz. Deposition was performed at a 650°C substrate temperature and under 10 mTorr oxygen pressure, followed by a post-annealing in oxygen rich environment (200 Torr) at 520°C for 30 min.

### Device fabrication

Optical waveguides were fabricated by electron beam lithography (EBL, Ellionix ELS7700) followed by HF etching. First a 30 nm thick W layer was deposited on the BTO sample, followed by spin coating of 300 nm thick resist (ma-N 2403). Then waveguides were patterned by EBL (50 keV) followed by development with a developer (ma-D 525). Subsequently, the W layer was etched by inductive coupled plasma (ICP, Oxford Instruments PlasmaPro 100 Cobra, process gas: $SF_6$/Ar, HF power: 40W, ICP power: 1000 W). The waveguides in BTO thin film was etched by diluted HF solution (0.03%). Finally, the end facets of waveguides were diced (Accretech SS20) and polished by focus ion beam (FIB, TESCAN GAIA3) to improve the fiber-chip coupling.

### Waveguide measurement

The measurement was carried out with a supercontinuum light source Amonics ALS-CL-15-B-FA and optical spectrum analyzer Agilent 86142B. The voltage is supplied by Keithley 2400.

### Simulation

Cross sectional mode analyses and electric field distribution simulations were performed in COMSOL Multiphysics. The MZI transmission spectra were simulated in Mathematica.

## Data availability

All data are available from the corresponding authors on reasonable request.

## Additional information

**Supplementary information.** The online version contains supplementary material available at https://doi.org/xxxxxx

**Correspondence and requests for materials** should be addressed to Yu Cao and Aaron Danner.

# Giant electro-optic coefficient in single-crystal barium titanate-on-oxide insulator based Mach-Zehnder interferometer

In the format provided by the authors and unedited

## Supplementary note S1

### Extraction of Pockels coefficient $r_{42}$

To extract the Pockels coefficient $r_{42}$ of the barium titanate (BTO) thin film from the measured transmission spectrum for the Mach-Zehnder interferometer (MZI), we performed theoretical analysis as follows.

Firstly, we performed cross-sectional analysis in COMSOL, as shown in Fig. 2 in the main text. We simulated the optical eigenmodes, and obtained the effective mode index of the fundamental TE mode. The corresponding optical power confinement $\Gamma_{\mathrm{BTO}}$ in the active BTO layer is given by

$$\Gamma_{\mathrm{BTO}} = \frac{|\int_{A_{BTO}} S\, ds|}{|\int_{A_{\infty}} S\, ds|} \tag{1}$$

where $S$ is the Poynting vector, and $A_{BTO}$ and $A_{\infty}$ represent the cross sectional areas of BTO and all space, respectively. We then simulated the DC electric field strength $E$ across the waveguide area where the optical mode was distributed.

With the results from cross-sectional simulations, we theoretically calculated the interference spectrum for the MZI device under an externally applied electric field.

Under the Pockels effect, with applied electric field strength $E$, the relative permittivity $\boldsymbol{\varepsilon}$ of BTO becomes

$$\boldsymbol{\varepsilon} = \varepsilon_0 \begin{bmatrix} n_o^2 & 0 & 0 \\ 0 & n_o^2 & -\Gamma_{\mathrm{BTO}}\,(n_o^2 n_e^2\, r_{42} E) \\ 0 & -\Gamma_{\mathrm{BTO}}\,(n_o^2 n_e^2\, r_{42} E) & n_e^2 \end{bmatrix} \tag{2}$$

where $n_o$ and $n_e$ are the ordinary and extraordinary indices.

If light is propagating in the *x* direction, the permittivity matrix can be reduced to

$$\boldsymbol{\varepsilon} = \varepsilon_0 \begin{bmatrix} n_o^2 & -\Gamma_{\mathrm{BTO}}\,(n_o^2 n_e^2\, r_{42} E) \\ -\Gamma_{\mathrm{BTO}}\,(n_o^2 n_e^2\, r_{42} E) & n_e^2 \end{bmatrix} \tag{3}$$

corresponding to a rotation of optic axis by an angle of $\theta$, which can be represented by a rotation matrix $\boldsymbol{R}$

$$\boldsymbol{R} = \begin{bmatrix} \cos\theta & -\sin\theta \\ \sin\theta & \cos\theta \end{bmatrix}. \tag{4}$$

Using this rotation matrix, the index matrix $\boldsymbol{\varepsilon}$ can be diagonalized to

$$\boldsymbol{\varepsilon}' = \varepsilon_0 \begin{bmatrix} n_a^2 & 0 \\ 0 & n_b^2 \end{bmatrix} \quad (5)$$

where $\boldsymbol{\varepsilon} = \boldsymbol{R}^{-1}\boldsymbol{\varepsilon}'\boldsymbol{R}$, and $n_a$ and $n_b$ represent the refractive indices along the new optic axes.

Assuming the optical field strength at input of the BTO MZI is normalized to 1, the optical field for the TE mode at the output can be represented as

$$E_x = cos^2\,\theta\, e^{-\frac{i2\pi n_a L}{\lambda}} + sin^2\,\theta\, e^{-\frac{i2\pi n_b L}{\lambda}} + e^{-i2\pi n_o (L+D)/\lambda} \quad (6)$$

where $L$ and $D$ are the electrode length and arm length difference respectively. The power of TE at the output is $P = E_x^2$.

The simulated resonance shift for our BTO MZI with an applied electric field of 5 V is shown in Fig. 2e. By matching the resonance shift with the measured spectrum for the MZI as shown in Fig. 2c, we extracted an $r_{42}$ of 1268 pm/V.